\documentclass[10pt,twocolumn,letterpaper]{article}

\usepackage{cvpr}              
\usepackage{url}
\usepackage[accsupp]{axessibility}
\usepackage[dvipsnames]{xcolor}

\usepackage{amssymb}
\definecolor{cvprblue}{rgb}{0.21,0.49,0.74}
\usepackage[pagebackref,breaklinks,colorlinks,citecolor=cvprblue]{hyperref}

\def\paperID{6} 
\def\confName{CVPR}
\def\confYear{DCAMI 2024}

\title{\lowercase{nn}M\lowercase{obile}N\lowercase{et}: Rethinking CNN for Retinopathy Research}

\author{ Wenhui Zhu$^{1}$\thanks{\textit{These authors contributed equally to this paper}} \enspace Peijie Qiu$^{2}$\footnotemark[1] \enspace Xiwen Chen$^{3}$\footnotemark[1]  \enspace Xin Li$^{1}$ \enspace Natasha Lepore$^{4}$ \enspace  Oana M. Dumitrascu$^{5}$ \\
\enspace Yalin Wang$^{1}$\thanks{\textit{Corresponding Author}}\\
$^{1}$ School of Computing and Augmented Intelligence, Arizona State University\\ 
$^{2}$ McKeley School of Engineering, Washington University in St. Louis\\ 
$^{3}$ School of Computing, Clemson University\\ 
$^{4}$ CIBORG Lab, Department of Radiology Children’s Hospital Los Angeles\\ 
$^{5}$ Department of Neurology, Mayo Clinic\\
{\tt\small wz52@asu.edu, ylwang@asu.edu }
}

\begin{document}
\maketitle

\begin{abstract}
Over the past few decades, convolutional neural networks (CNNs) have been at the forefront of the detection and tracking of various retinal diseases (RD). Despite their success, the emergence of vision transformers (ViT) in the 2020s has shifted the trajectory of RD model development. The leading-edge performance of ViT-based models in RD can be largely credited to their scalability—their ability to improve as more parameters are added. As a result, ViT-based models tend to outshine traditional CNNs in RD applications, albeit at the cost of increased data and computational demands. ViTs also differ from CNNs in their approach to processing images, working with patches rather than local regions, which can complicate the precise localization of small, variably presented lesions in RD. In our study, we revisited and updated the architecture of a CNN model, specifically MobileNet, to enhance its utility in RD diagnostics. We found that an optimized MobileNet, through selective modifications, can surpass ViT-based models in various RD benchmarks, including diabetic retinopathy grading, detection of multiple fundus diseases, and classification of diabetic macular edema. The code is available at \url{https://github.com/Retinal-Research/NN-MOBILENET}
\end{abstract}


\section{Introduction}
\label{fig:main}
Retinal diseases (RD), such as diabetic retinopathy (DR), age-related macular degeneration, inherited retinal conditions, myopic maculopathy, and retinopathy of prematurity, are major contributors to blindness worldwide~\cite{retinaldisease}. Deep neural networks, particularly convolutional neural networks (CNNs), have been extensively used in retinal image analysis over the past decades, achieving cutting-edge results in various RD-related tasks~\cite{denosing1,denosing2,CANet,zoom-in-net,sem+adv,AFN,comp-expertab,DETACH,selfmil}. The effectiveness of CNNs in these applications is largely due to their built-in architectural inductive biases, such as spatial hierarchies, locality, and translation invariance. These characteristics enable CNNs to transform local visual elements like edges and textures into complex, high-level abstracted features. Building on this approach, numerous CNN-based RD models~\cite{zoom-in-net,sem+adv,CANet,DETACH} have incorporated disease-specific biases into their designs. However, the specialized nature of these CNN-based models for RD limits their versatility across a range of RD tasks.

\begin{figure}[!t]
    \centering
    \includegraphics[width=1.0\columnwidth]{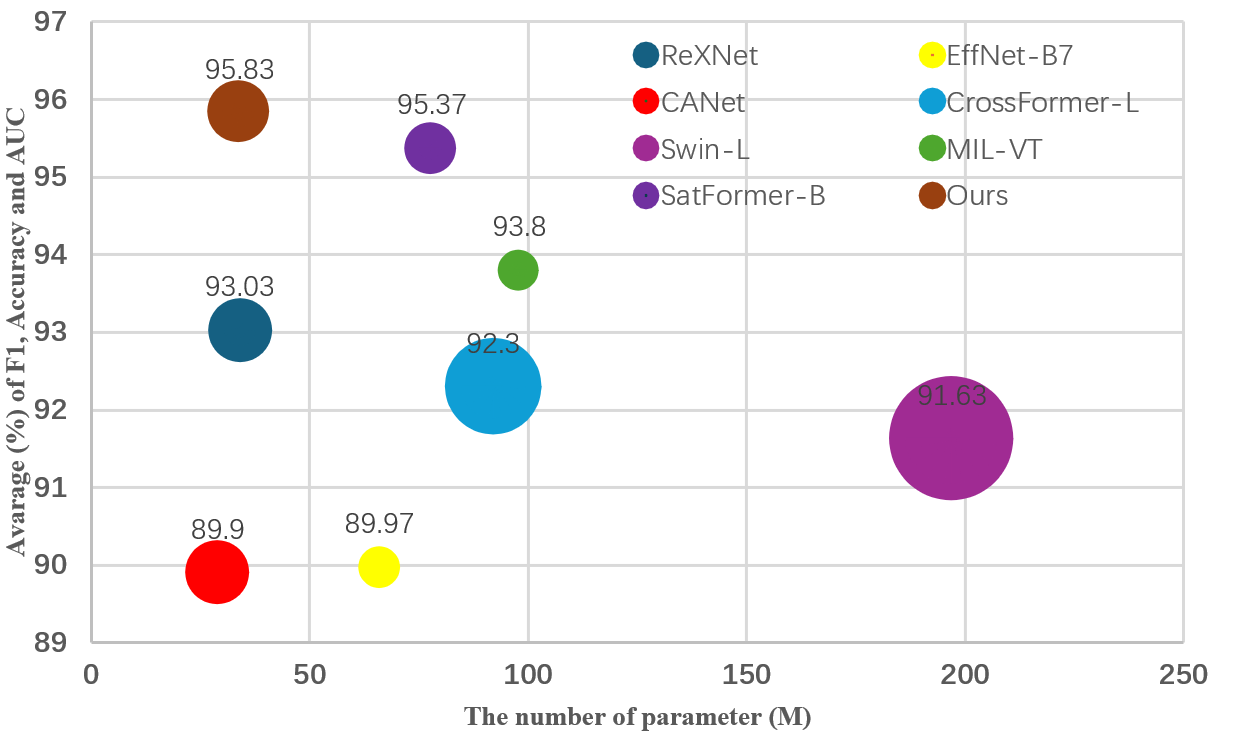}
    \caption{Model size vs average performance (F1, Accuracy and AUC) on retinal multi-disease abnormal detection using RFMid dataset. Our method demonstrates superiority over other CNN/ViT based methods in terms of performance and efficiency. }
    \label{fig:performance}
\end{figure}

Recent advancements in RD modeling~\cite{MIL-VT,SatFormer-B,LAT,swin-L,crossformer} have largely revolved around the vision transformer~\cite{dosovitskiy2020image} (ViT) since its debut in the 2020s. The prowess of ViT-based models in RD is primarily due to their capacity to scale effectively; their performance improves as the model size grows~\cite{liu2022convnet}. Consequently, ViT-based methods typically exhibit superior performance over CNNs but with a cost of computational burden (as evident in Fig.~\ref{fig:performance}). In addition, ViT-based models are advantageous for capturing long-range global dependencies via  self-attention mechanism. Nonetheless, the quadratic time and memory complexity of self-attention operation make ViTs computationally intense and data-hungry. Accordingly, ViT-based RD models typically necessitate pretraining on large-scale datasets~\cite{MIL-VT,SatFormer-B,swin-L}. Furthermore, unlike CNNs, these models generally lack locality, since they operate at the image patch level~\cite{swin-L}.

To address these challenges, new iterations of ViT designs bring back convolution-like features to recover local context sensitivity~\cite{swin-L,liu2021swin}. This adaptation is particularly beneficial for RD
research since RD lesions are typically small and have heterogeneous appearances. One representative example can be found in the DR task, where four distinct types of lesions (i.e., microaneurysms, soft exudates, hemorrhages, and hard exudates) exhibit variations in shape, size, structure, and contrast. Among these lesions, the microaneurysms are too tiny to be easily detected. In addition, the DR grading task inherently contains hierarchical information, e.g., a proliferative DR image may consist of all types of lesions. These intrinsic properties in RD tasks naturally align with inductive biases in CNNs, where hierarchical and fine-grained local contexts can be better detected than ViTs. This observation prompts a reconsideration: \emph{could CNNs be inherently more suited to RD tasks than ViTs?}

We also note that recent studies have shown well-tuned CNNs surpassing ViTs in general image classification tasks~\cite{liu2022convnet}. Motivated by these findings, our research recalibrated a CNN, specifically MobileNetV2~\cite{sandler2018mobilenetv2}, for RD applications, focusing on training strategies and architectural refinements such as the inverted bottleneck, dropout optimization, and activation functions. To this end, we introduce nnMobileNet (``no-new" MobileNet), a model that implements strategic yet minimal enhancements. Our empirical results confirmed nnMobileNet's superiority over many leading RD models across a spectrum of benchmarks (see Fig.~\ref{fig:performance}). Our work not only substantiates the potential of CNNs in RD research but also emphasizes the critical aspects of CNN optimization. 
We anticipate that our findings will spark a renewed interest in the adaptability and fine-tuning of CNNs within the field.

\section{Related Works}
\subsection{Diabetic Retinopathy Assessment} In the domain of deep learning, the evaluation of diabetic retinopathy (DR) encompasses two tasks: DR grading and the classification of referable DR. The process of DR grading adheres to a protocol that categorizes the progression of diabetic retinopathy into distinct stages based on lesion examination, facilitating a multi-class classification task. This grading delineates five levels of severity: no retinopathy, mild non-proliferative DR (NPDR), moderate NPDR, severe NPDR, and proliferative DR (PDR). In contrast, the task of identifying referable DR focuses on detecting that may lead to blindness or significant vision loss resulting from DR, thereby being treated as a binary classification framework. The framework distinguishes between non-referable DR, characterized by the absence or mild presence of NPDR without significant pathological manifestations, and referable DR (rDR), which encompasses conditions of moderate severity or higher. In the past decade, deep learning has achieved state-of-the-art performance in automating the diagnosis of DR. Following the trend, convolutional neural networks (CNN) dominated the early stage of development~\cite{CANet,zoom-in-net,sem+adv,AFN,DETACH,comp-expertab,MIL-VT}. Among them, Zoom-in-Net~\cite{zoom-in-net} took a biomimetic method (medical experts utilized the magnification to locate the lesion in the diagnosis) that incorporated the multiple scale information into CNN. Zhou et al. proposed a semi-supervised learning framework, which coordinated lesion segmentation and classification tasks by utilizing pixel-level supervision~\cite{sem+adv}. CANet~\cite{CANet} integrated two attention modules to jointly generate disease-specific and disease-dependent features for grading DR and diabetic macular edema (DME). Che et al.~\cite{DETACH} achieved good performance via robust disentangled features of DR/DME. Essentially,
These CNN-based DR classification methods rely on the extra auxiliary task and prior knowledge, which inevitably introduce more complex models and specialized multi-task datasets(e.g., DME Classification and lesion segmentation). Vision Transformers (ViT) have recently gained much attention in various visual tasks by leveraging the self-attention mechanism to capture long-term feature dependencies. Along this direction, MIL-VT~\cite{MIL-VT} proposed using multiple-instance pooling to aggregate the features extracted by a ViT. Sun et al.~\cite{LAT} proposed a lesion-aware transformer (LAT) to learn the diabetic lesion-specific features via a cross-attention mechanism. Although those methods achieved state-of-the-art performance, most heavily relied on pretraining on large-scale datasets due to the data-hungry nature of ViT whose complexity quadratically grew concerning the input size. In addition, the DR features are localized in nature, e.g..fine-grained lesions such as microaneurysms typically occupy only a minor fraction of the image area and are discretely distributed in vessels. It was challenging for pure transformer-based feature extractors to focus more on global representations.  In this paper, we contend that the capabilities of CNNs for DR tasks remain underutilized. We argue that through fine-tuning techniques, CNNs can achieve significant performance improvements, potentially surpassing ViT. To substantiate our claim, we initiate our investigation with a lightweight framework, MobileNet, and conduct a series of empirical studies.

\subsection{Multi Retinopathy abnormal detection} 
The Multi-Retinopathy delineates a broader subclassification of Retinopathy, introducing more precise representations of lesions. Several fundus images may carry one or multiple labels, such as asteroid hyalosis, anterior ischemic optic neuropathy, age-related macular degeneration, branch retinal vein occlusion, Choroidal folds, etc. Notably, many of these pathological changes are interrelated; for instance, the presence of cotton wool spots on the retina is a characteristic ocular manifestation of various medical conditions, including diabetes mellitus, systemic hypertension, leukemia, and AIDS~\cite{cotton}.  CNNs remain dominant as the foundational design approach for multi-retinopathy abnormal detection. A significant portion of the benchmark methods based on CNNs originates from DR grading models. A notable example of such work is the development of CANet~\cite{CANet}, which leverages multi-task learning to extract additional semantic information, thereby aiding the classification model. Most subsequent advancements in CNN-based methods have followed this conceptual framework~\cite{DETACH,CANet}.  Contrasting with this trend, some studies argue that establishing long-range dependencies and capturing global semantic information learning is a potentially more effective strategy for advancing model capabilities. The MIL-VT introduces the ViT and incorporates multiple instance learning head to force the token to capture the lesion information~\cite{MIL-VT}. However, this method processes each individual patch without emphasizing the semantics of smaller lesions, resulting in a lack of localized information modeling. Furthermore, they employed extensive external datasets for pre-training due to data-hangry nature of ViT.  In contrast, SatFormer enhances the ViT framework by integrating multi-scale CNNs to detect small lesions, such as microaneurysms and exudates. This approach enriches the model's capability to represent features of small lesions and to capture a wide range of pathological semantics~\cite{SatFormer-B}.  This transition and amalgamation from ViT back to CNN prompt us to ponder whether CNNs are more suited for RD than ViTs or whether the potential of CNNs remains underexploited. This curiosity underpins our motivation for conducting deeper research into the CNNs in various RD tasks.

\subsection{Myopic maculopathy grading} 
Recent trends show a growing interest in leveraging deep learning for the automatic diagnosis and analysis of myopic macular degeneration, the most prevalent complication of myopia and the leading cause of vision loss in individuals with pathological myopia.
At the recently concluded MICCAI 2023, the Automated Detection of Myopic Maculopathy in MMAC 2023 challenge featured tasks in myopic maculopathy grading, segmentation, and prediction of spherical equivalent. Insights from the released solutions reveal that the first-place winner utilized a two-stage pre-training method with a CNN backbone, incorporating vision-language pre-training and self-supervised visual representation learning. The second-place team employed a Swin Transformer backbone combined with ArcFace loss. Interestingly, the third-place entry achieved commendable results using a lightweight CNN model without needing external retinal datasets for pre-training or self-supervised learning~\cite{mmac1,mmac2,mmac3}. This outcome suggests that CNNs can still excel in performance, potentially outpacing ViTs in RD tasks. Moreover, the equitable testing environment of such challenges lends credibility to the results~\cite{why}. These findings corroborate our initial hypothesis and further pique our interest in exploring the fine-tuning of CNNs for RD tasks.



\section{Roadmap of a nnMobileNet}
\label{method}

\begin{figure}[!t]
    \centering
    \includegraphics[width=1.0\columnwidth]{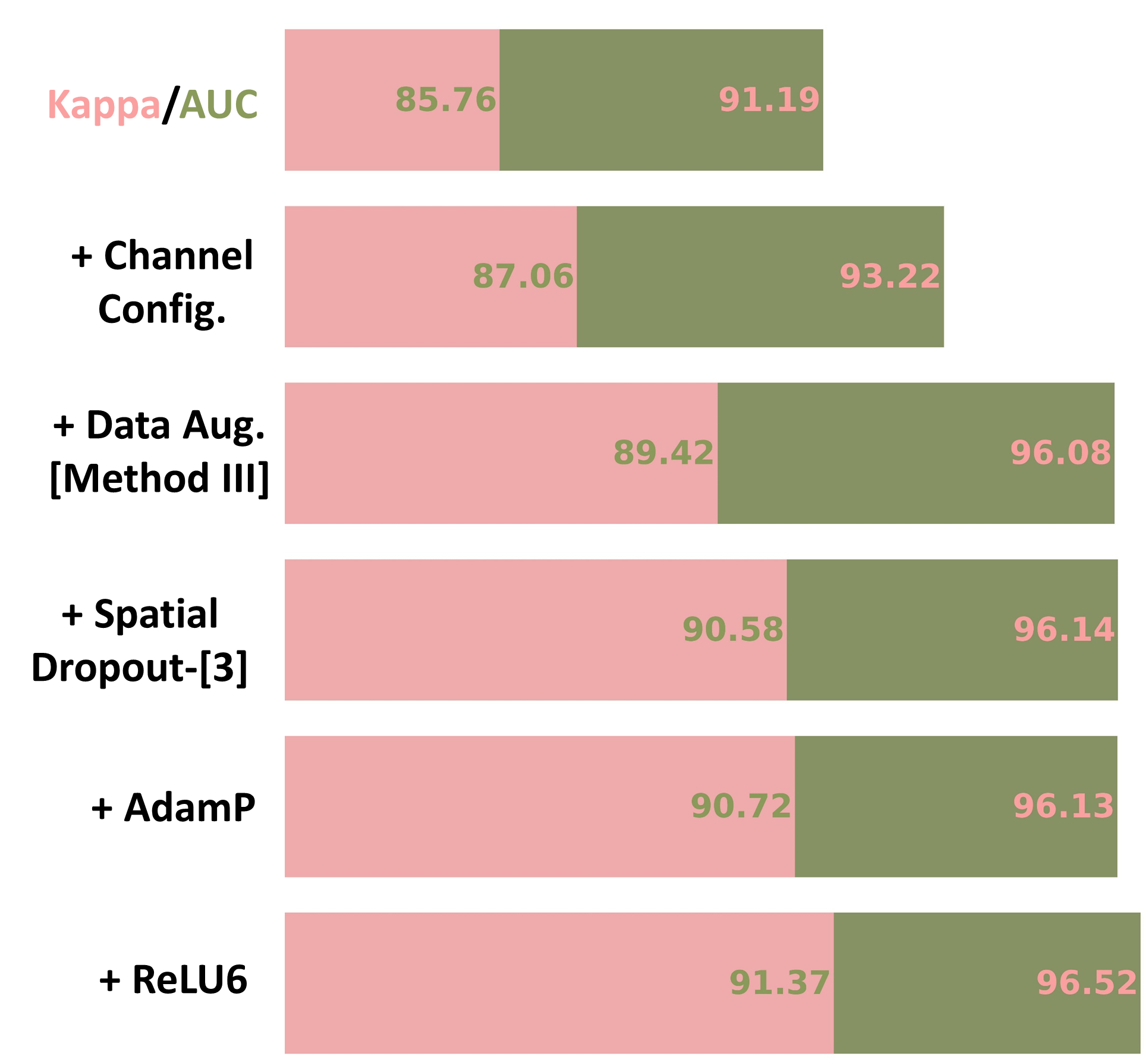}
    \caption{The roadmap of modifying a MobileNetV2 to the proposed no-new MobileNet (nnMobileNet) on the Messidor-2 dataset; }
    \label{fig:roadmap}
\end{figure}

\begin{figure*}[!t]
    \centering
    \includegraphics[width=0.8\textwidth]{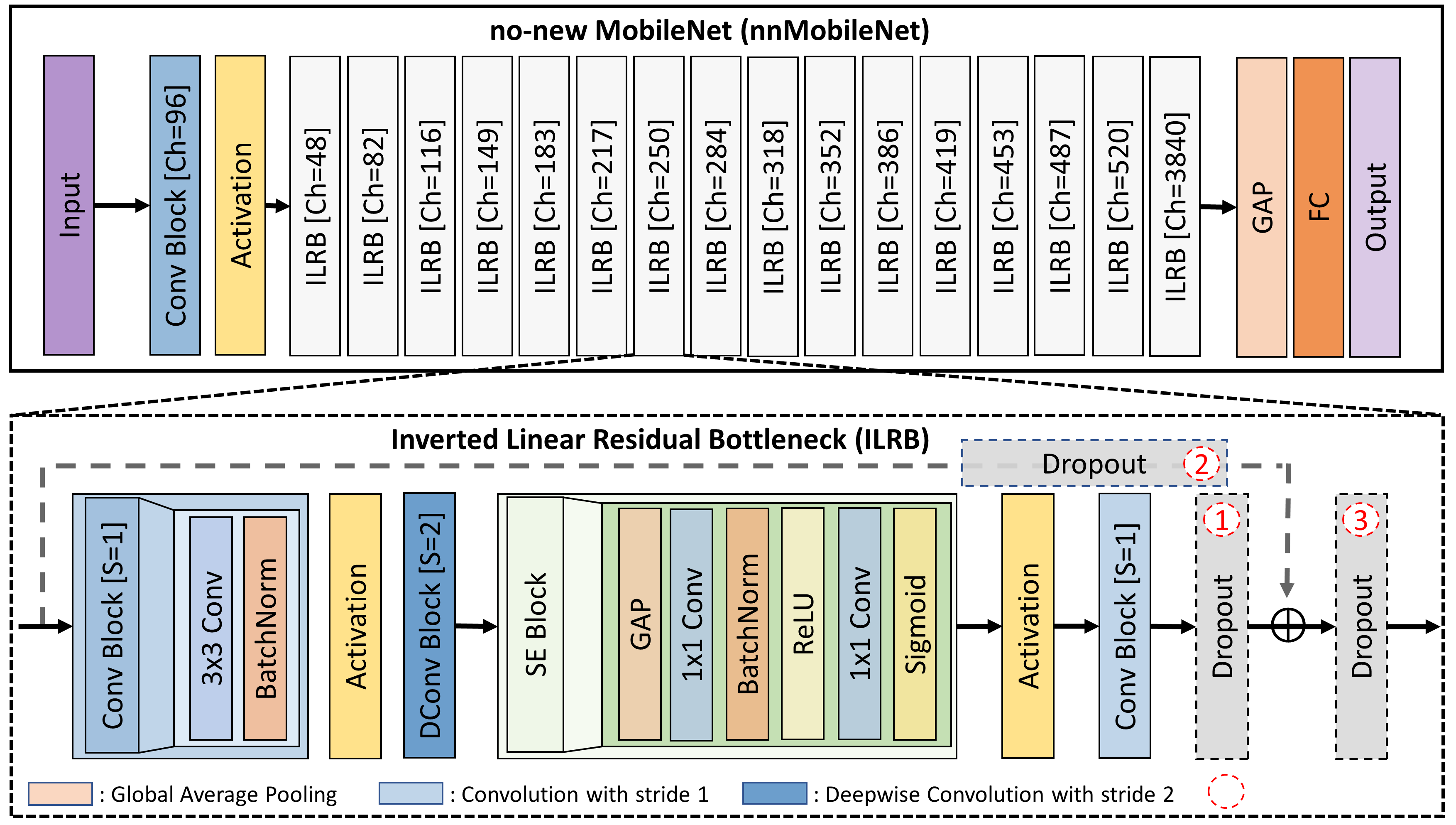}
    \caption{The detailed architecture of the no-new MobileNet (Including the Channel configuration) and the inverted linear residual bottleneck used in the no-new MobileNet.}
    \label{fig:network}
\end{figure*}

\begin{figure}[!t]
    \centering
    \includegraphics[width=0.9\columnwidth]{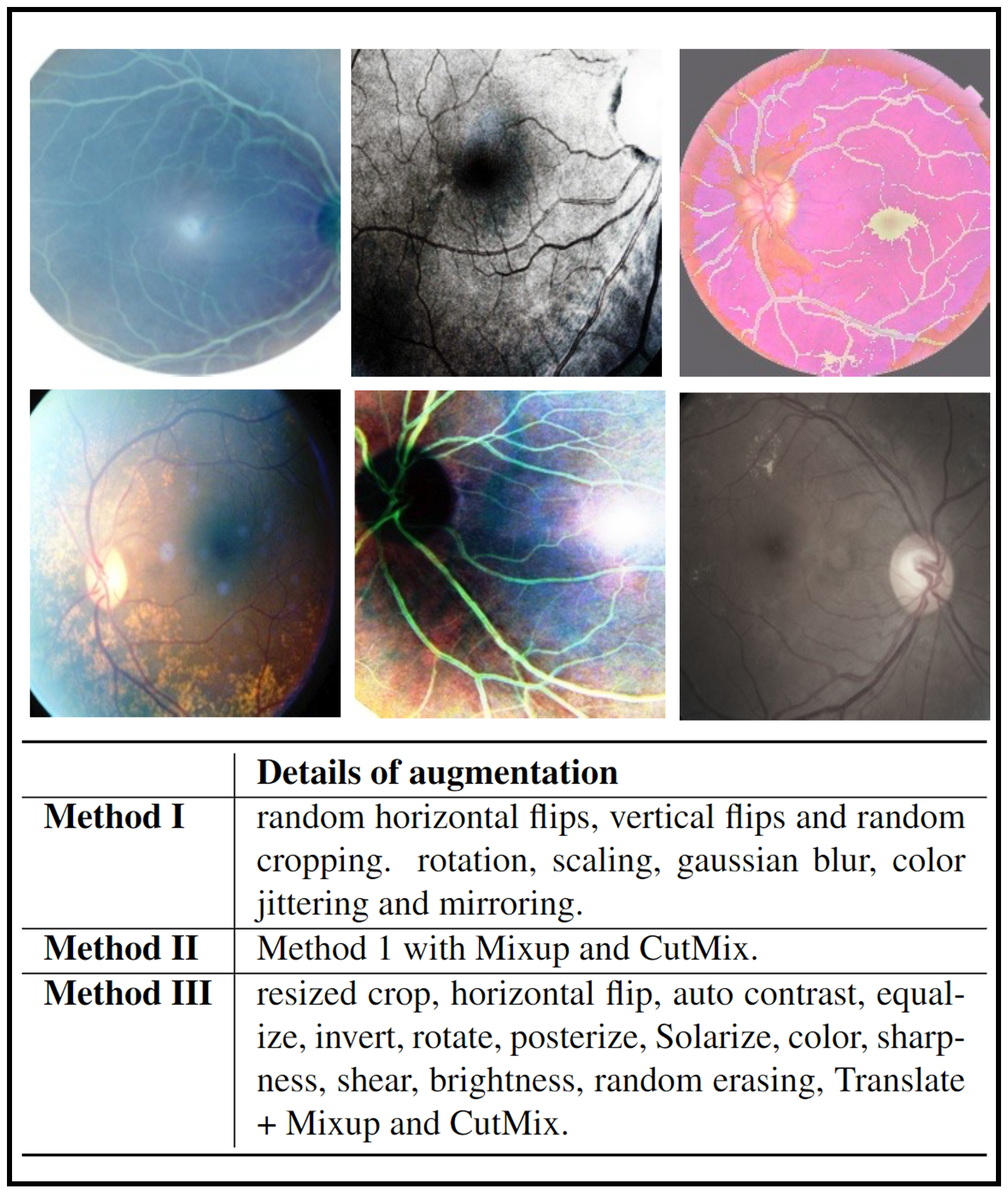}
    \caption{Examples of data augmentation (Method \uppercase\expandafter{\romannumeral3}) and details of three sets of data augmentation we used. }
    \label{fig:da}
\end{figure}

\label{sec:roadmap}
Our investigation started with a standard MoblieNetV2~\cite{sandler2018mobilenetv2} on the Messidor-2 dataset (see the first row in Fig.\ref{fig:roadmap})~\cite{mesidor}. 
We chose MobileNetV2 because of its efficiency, achieved by replacing the traditional residual bottleneck~\cite{resnet} with an inverted linear residual bottleneck (ILRB) where channel attention was included by default (see details in Fig.\ref{fig:network}).
We conducted empirical studies on its key components, including channel configuration, data augmentation, dropout, optimizer, and activation functions. 

\subsection{Channel Configuration of ILRB}
\label{sec:CC}
Recent findings in natural images have revealed that changing stage-wise channel configuration (primarily computation distribution between layers) led to a remarkable performance gain~\cite{ReXNet,liu2022convnet}. This improvement is primarily due to two factors: first, the expressiveness of a layer is affected by the rank of its output matrix~\cite{ReXNet}. Second, ViTs generally use a different stage ratio from CNNs to change its computation distribution~\cite{liu2022convnet}. Both of these factors indicated the necessity of changing the channel configuration in MobileNetV2.

As consistent with findings in natural images, we empirically found that changing the channel configuration led to a performance gain of 2.03\% in Kappa and 1.30\% in AUC  (see the second row in Fig.~\ref{fig:roadmap}). It is worth noting that we follow the channel configuration in~\cite{ReXNet} (see Fig.\ref{fig:network}  no-new MobileNet architecture for channel configuration details).   

\subsection{Data Augmentation}
\label{sec:augmentation}
In the field of RD, a common belief was that heavy data augmentation (e.g., Mixup and CutMix) should be avoided, as it dramatically distorts structures of images~\cite{LAT,SatFormer-B,CANet}. However, recent research has revealed that heavy data augmentation even boosted the performance in retinal vessel segmentation~\cite{vessel}. We hypothesize that this finding can be transferred to the RD tasks because introducing noise from the heavy data augmentation (e.g., unrealistic images shown in Fig.\ref{fig:da}) may help the model generalize better.
To validate those hypotheses, we conducted experiments on three different sets of data augmentation combinations from light to heavy (as detailed by Methods \uppercase\expandafter{\romannumeral1}, \uppercase\expandafter{\romannumeral2}, \uppercase\expandafter{\romannumeral3} in Fig.\ref{fig:da}). 

Interestingly, we found that the heaviest data augmentation (i.e., Method III) achieved the best performance compared to the other two strategies (see Fig.~\ref{fig:main}). We conjectured that different from ViTs which perform classification by comparing patches, the lack of non-local context in CNNs may necessitate heavy data augmentation to find the most discriminative local feature representation for abstracting heterogeneous RD lesions. Integrating this set of data augmentation into our nnMobileNet model design can improve Kappa by 2.36\% and AUC by 2.86\% (see the third row in Fig.~\ref{fig:roadmap}). 


\begin{figure}[!t]
    \centering
    \includegraphics[width=0.99\columnwidth]{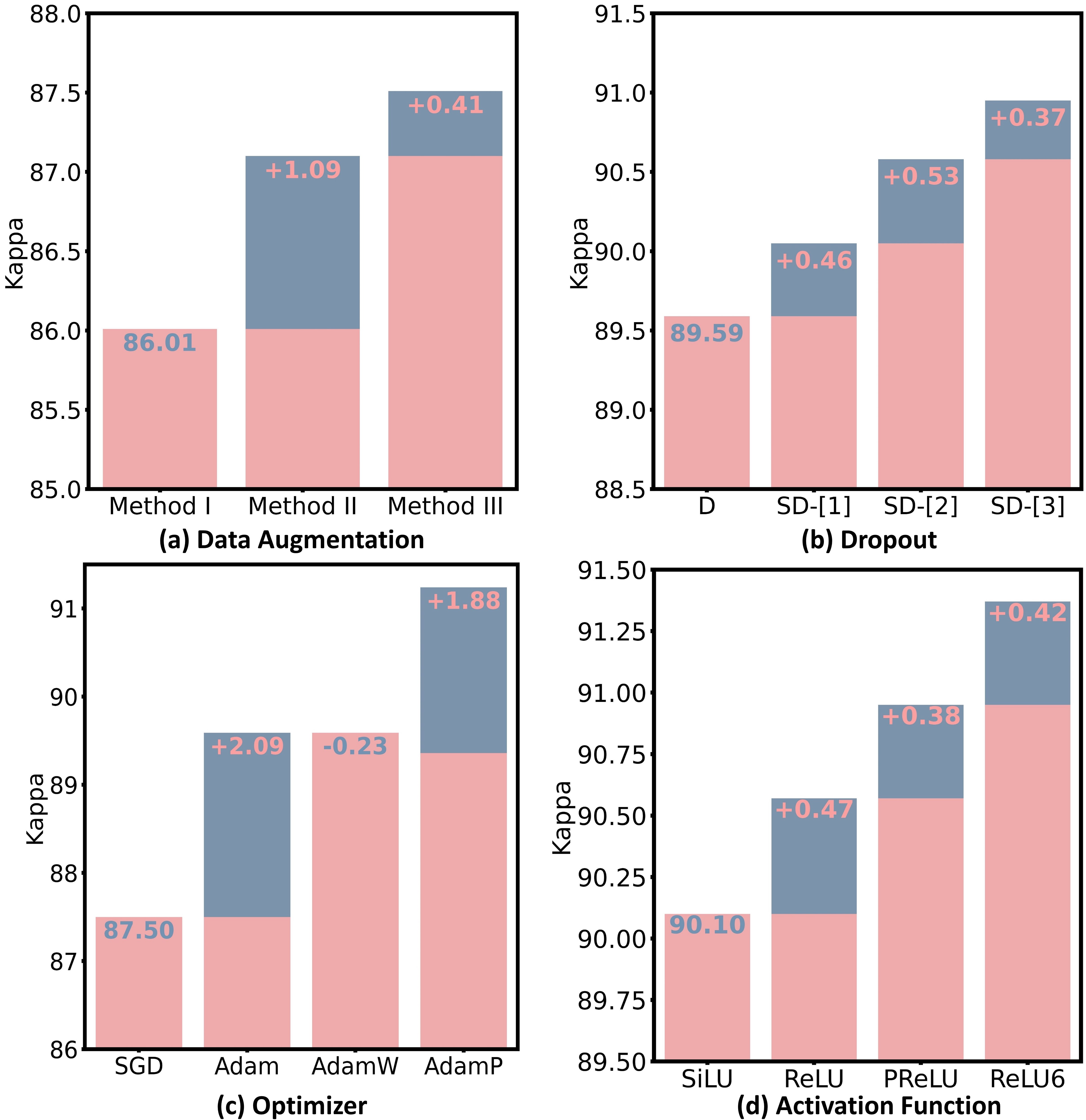}
    \caption{Empirical studies on Messidor-2 dataset where subpanel pictures (a), (b), (c), and (d) represent different experimental groups, each of which is independent of the others. D and SD-[x] in subpanel (b) denote Dropout and SpatialDropout in position [x] as shown in Fig.\ref{fig:network}(c), respectively.}
    \label{fig:main}
\end{figure}

\subsection{Dropout}
\label{sec:Dropout}
Dropout is commonly employed to alleviate overfitting and enhance the representational capacity of individual layers. 
However, we argue that the standard Dropout~\cite{dropout}, which randomly zeros out neurons in a feature map, is unsuitable for general RD tasks. This is mainly due to the fact that specific lesion information is primarily encoded in certain image color channels. An example can be seen in the case of DR, where the information of microaneurysm predominantly resides in the red channel, while the exudate information is primarily encoded in the green channel. Based on this observation, we conjecture that spatial Dropout, which randomly removes channels from a feature map, is a better choice for RD tasks. Additionally, spatial Dropout naturally preserves spatial and local structures by randomly dropping out strongly activated local patterns~\cite{dropout_chanelwise}, which is highly needed in RD. However, \emph{where to place spatial Dropout remains an open problem.} Here, we conducted experiments to investigate the strategic placement of spatial Dropout in ILRB (see Fig.\ref{fig:network} inverted linear residual bottleneck). 

We observed that i) spatial Dropout is indeed more effective than standard Dropout for RD tasks and ii) the performance varies when placing the spatial Dropout at different positions (see Fig.\ref{fig:main}(b)). 
Integrating spatial Dropout into the proposed model led to an improvement of 0.06\% in Kappa and 1.16\% in AUC (see the forth row in Fig.\ref{fig:roadmap}).

\subsection{Optimizer}
\label{ssec:optimizer}

The training of ViTs is typically performed by an AdamW~\cite{adamw} optimizer, which makes us wonder if the performance gain in ViT-based RD models comes from the more advanced optimizer~\cite{liu2022convnet}. Alternatively, \emph{would a more advanced optimizer boost the performance of a CNN-based RD model ?} 

Our empirical studies revealed that training the network with AdamP~\cite{adamp} optimizer, which better accommodates the step size adaptively, showed better performance compared to other optimizers (see Figure~\ref{fig:main}(c)). Applying the AdamP to train nnMobileNet contributed to a performance gain of 2.2\% in Kappa (see the fifth row in Fig.~\ref{fig:roadmap}).

\subsection{Activation Function}
\label{ssec:AF}
ReLU is extensively used in traditional CNNs due to its simplicity and computational efficiency. However, increasing research indicates that smoother variants of ReLU (e.g., SiLU), commonly used in ViTs, can lead to performance improvements~\cite{liu2022convnet,ReXNet}. Based on that, we investigate the most suitable activation functions within inverted linear residual blocks (see Fig.\ref{fig:network} inverted linear residual bottleneck) for retinal disease applications. Here, we consider four variants of ReLU, including SiLU, ReLU, PReLU, and ReLU6. As shown in Fig.\ref{fig:main}(d), the ReLU6 activation was the best among all options. After we replaced the ReLU with ReLU6 in each ILRB, it led to an improvement of 0.65\% in kappa and 0.39\% in AUC (see the sixth row in Fig.~\ref{fig:roadmap}).

\section{Experiments and Results}

An optimal set of network structures and training strategies is summarized in Section~\ref{method}. We used cross-entropy loss for training all the models in this work. All models were trained for 1000 epochs with a batch size of 32.
The initial learning rate was set to 0.001 decayed according to a cosine decay learning rate scheduler with 20 epochs of linear warm-up. A weight decay rate of 0.05 was applied to prevent overfitting. All experiments were implemented in PyTorch and were performed on a Nvidia RTX 3090 GPU with a memory of 24G.

\subsection{Datasets and Evaluation Metrics}
\noindent \textbf{Messidor-1 dataset}~\cite{mesidor} contains 1200 fundus images with four DR grades. We conducted referral and normal DR classification in this dataset. In the referral and non-referral DR classification, Grades 0 and 1 are considered non-referable, while Grades 2 and 3 are considered referable DR (rDR). For normal and abnormal classification, only Grade 0 will be labeled as normal, and the other grades will be recognized as abnormal. 
We followed the experimental settings in~\cite{CANet} by using 10-fold cross-validation on the entire dataset. The area under the curve (AUC) was used as the evaluation metric. 

\noindent \textbf{Messidor-2 dataset}~\cite{mesidor} contains 1748 fundus images with five DR grades. As no official split of the training and testing dataset was provided, we used this dataset to conduct ablation studies to demonstrate the effectiveness of each component of our proposed method on DR grading evaluated by the AUC and quadratic Cohen's kappa (Kappa). 

\noindent \textbf{RFMiD dataset}~\cite{rfmiddataset} contains 1920 training, 640 validation, and 640 testing images with 45 different types of pathologies (central serous retinopathy, central retinal vein occlusion, asteroid hyalinosis, etc.). Following the protocol in \cite{SatFormer-B,MIL-VT}, we performed normal and abnormal binary classification on this dataset whose performance is measured by accuracy (ACC), AUC, and F1. 

\noindent \textbf{APOTS dataset}~\cite{APOTS} contains 3662 fundus images for DR grading with the severity on a grade of 0 to 4 (no DR, mild, moderate, severe, proliferative DR). Following the experimental setting of 5-fold cross-validation in \cite{MIL-VT}, we evaluated the performance of DR grading in terms of ACC, AUC, weighted F1, and kappa. 

\noindent \textbf{IDRiD dataset}~\cite{idrod} contains 413 training and 103 testing images for both DR grading and DME severity grading tasks. we used the training and testing data provided by the official split. Different from method~\cite{CANet} that re-labeled DR grading into two categories, we trained the multi-class DR grading task and reported the evaluation metrics of ACC, AUC, and F1. Both grading experiments followed the protocol in~\cite{DETACH}. 

\noindent \textbf{MICCAI 2023 MMAC (Myopic Maculopathy Analysis Challenge)} contains 1143 fundus images with four myopic maculopathy grades. There are 404 images for grade 0, 412 images for grade 1, 224 images for grade 2,60 images for grade 3, and 43 images for grade 4. We used 5-fold stratified cross-validation on the training set. The Quadratic-weighted Kappa (kappa), F1 score, and Specificity were used as the evaluation metric. For this experiment, we felt the raw data was good quality and did not need to apply any preprocessing.


\subsection{Comparison to State-of-the-art Methods}

\begin{table}[!t]
    \centering
     \caption{Comparison of rDR and normal classification on the Messidor-1 dataset~\cite{mesidor}. Annotations denote whether pixel-level or patch-level supervision was applied. ($^\dagger$: methods implemented by us; while the other benchmarks are taken from~\cite{vbxk-ckml,LAT,CANet}.)}
     \resizebox{0.45\textwidth}{!}{
    \begin{tabular}{lccc}
        \toprule
                \multicolumn{1}{c}{}  &  \multicolumn{3}{c}{}\\
         Method  &  Annotations & Referral AUC & Normal AUC \\
        \midrule
        VNXK \cite{vbxk-ckml}  &  - & 88.7 & 87.0  \\
        CKML \cite{vbxk-ckml}& -   &  89.1 &  86.2  \\
        Comp. CAD \cite{comp-expertab} &  - &  91.0  &  87.6 \\
        Expert A \cite{comp-expertab} &  - & 94.0 & 92.2 \\
        Expert B \cite{comp-expertab}&  - & 92.0 & 86.5 \\
        Zoom-in-Net  \cite{zoom-in-net} &  - & 95.7 & 92.1  \\
        AFN  \cite{AFN}&  patch & 96.8 & -  \\
        Semi + Adv   \cite{sem+adv} &  pixel & 97.6 & 94.3\\
        $^\dagger$CANet \cite{CANet} &  - & 96.3 & - \\
        LAT  \cite{LAT} &  - &  98.7 & 96.3  \\
        \cline{1-4} 
        Ours  & -  &    \textbf{98.7} &    \textbf{ 97.5} \\        
        \bottomrule
    \end{tabular}
}
    \label{tb:messidor}
\end{table}

\begin{table}[t]
    \centering
    \caption{Performance comparison of multi-disease abnormal detection on the RFMiD dataset~\cite{rfmiddataset}. Param are the parameter numbers, indicating model complexity of models. (Due to some methods codes not being made publicly available, $^\dagger$: methods reproduced by us; while the other benchmarks are taken from~\cite{SatFormer-B}.)}
    \begin{tabular}{lccc}
        \toprule
                \multicolumn{1}{c}{}    &  \multicolumn{3}{c}{ Normal/Abnormal}\\
             \cmidrule(lr){2-4} 
         Method    & ACC & AUC & F1 \\
        \midrule
        $^\dagger$ CANet \cite{CANet}  & 88.3 & 91.0  & 90.4 \\
        $^\dagger$ EffNet-B7 \cite{effNet-B7}   &  88.2 &  91.0  &  90.7 \\
        $^\dagger$ ReXNet \cite{ReXNet} &  91.3 &  94.5 &  93.3 \\
        $^\dagger$ CrossFormer-L\cite{crossformer} & 90.6 & 94.3 & 92.0 \\
        $^\dagger$ Swin-L \cite{swin-L}  & 89.5 & 93.8 & 91.6\\
        $^\dagger$ MIL-VT \cite{MIL-VT}  & 91.1 & 95.9 & 94.4 \\
        SatFormer-B \cite{SatFormer-B}  & 93.8 &  96.5 &  \textbf{95.8} \\
        \cline{1-4}
        Ours  & \textbf{94.4} &   \textbf{98.7} &  94.4\\        

        \bottomrule
    \end{tabular}
    
    \label{tb:RFMiD}
\end{table}

\begin{table}[t]
    \centering
    \caption{Performance comparison of DR grading on the APOTS dataset~\cite{APOTS}. Due to some methods codes not being made available in public, $^\dagger$ denotes methods reproduce performances at the same level as reported while the other benchmarks are taken from~\cite{MIL-VT}.}
    \resizebox{0.80\columnwidth}{!}{
    \begin{tabular}{lcccc}
    
        \toprule
                \multicolumn{1}{c}{}  &  \multicolumn{4}{c}{ DR Grading}\\
             \cmidrule(lr){2-5} 
         Method   & AUC & ACC & F1 & Kappa \\
        \midrule
        DLI   \cite{DLI}  &  - & 82.5 & 80.3  & 89.0 \\
         $^\dagger$ CANet  \cite{CANet}  & -   &  83.2 &  81.3  &  90.0 \\
        GREEN-ResNet50  \cite{green}  &  - &  84.4  &  83.6 &  90.8 \\
        GREEN-SE-ResNext50 \cite{green}  &  - & 85.7 & 85.2 & 91.2 \\
         $^\dagger$ MIL-VT \cite{MIL-VT}  &  \textbf{97.9} & 85.5 & 85.3 & 92.0 \\
        \cline{1-5}
        Ours  & 97.8  &    \textbf{89.1} &    \textbf{88.9} &   \textbf{93.4}\\        

        \bottomrule
    \end{tabular}
}
    \label{tb:APOTS2019}
\end{table}

\begin{table}[!t]
    \centering
    \caption{Performance comparison of DR and DME grading on the IDRID dataset~\cite{idrod}. $^\dagger$ denote methods implemented by us while the other benchmarks are taken from~\cite{DETACH}.}
    \resizebox{0.95\columnwidth}{!}{
    \begin{tabular}{lccc|ccc}
        \toprule
                \multicolumn{1}{c}{}  & \multicolumn{3}{c}{DME} & \multicolumn{3}{c}{DR}\\
             \cmidrule(lr){2-4}  \cmidrule(lr){5-7}   
         Method  & AUC& F1& ACC& AUC& F1& ACC \\
        \midrule
         $^\dagger$ CANet \cite{CANet} & 87.9 & 66.1 & 78.6      & 78.9 & 42.3 & 57.3       \\
        Multi-task net  \cite{Multitask-Net}& 86.1 & 60.3 & 74.8      & 78.0   & 43.9 & 59.2       \\
         $^\dagger$ MTMR-net \cite{MTMR-NET} & 84.2 & 61.1 & 79.6      & 79.7 & 45.3 & 60.2       \\
         $^\dagger$ DETACH + DAW ~\cite{DETACH} & 89.5 & 72.3 & 82.5      & 84.8 & 49.4 & 59.2       \\
        \cline{1-7}
        Ours & \textbf{95.3} & \textbf{84.8} & \textbf{86.5} & \textbf{91.6} & \textbf{72.6} & \textbf{73.1}      \\        

        \bottomrule
    \end{tabular}

    \label{tb:idrid}
    }
\end{table}

\begin{table}[!t]
    \centering
    \caption{The performance comparison in 2023 MMAC Challenge. The results are tested by the officially challenge platform \cite{challenge}.}
\resizebox{0.45\textwidth}{!}{    
    \begin{tabular}{l|ccccc}
        \toprule
         Method   & Kappa & F1 & SPE & Average & CPU time (s)\\
        \midrule
         Rank 1$^{st}$~\cite{mmac1}& 90.1 &  78.1 & 94.5 & 87.5 & 2.1283  \\
         Rank 2$^{nd}$~\cite{mmac2} & 88.9 &  76.8 & 94.1 & 86.6 & 0.8047  \\   \cline{1-6}  
         Ours (can reach 3$^{nd}$)& 90.0 &  75.1 & 94.1 & 86.4 & 0.2750  \\   

        \bottomrule
    \end{tabular}
}
    \label{tb:finalrank}
\end{table}

\begin{figure*}[!t]
    \centering
    \includegraphics[width=0.85\textwidth]{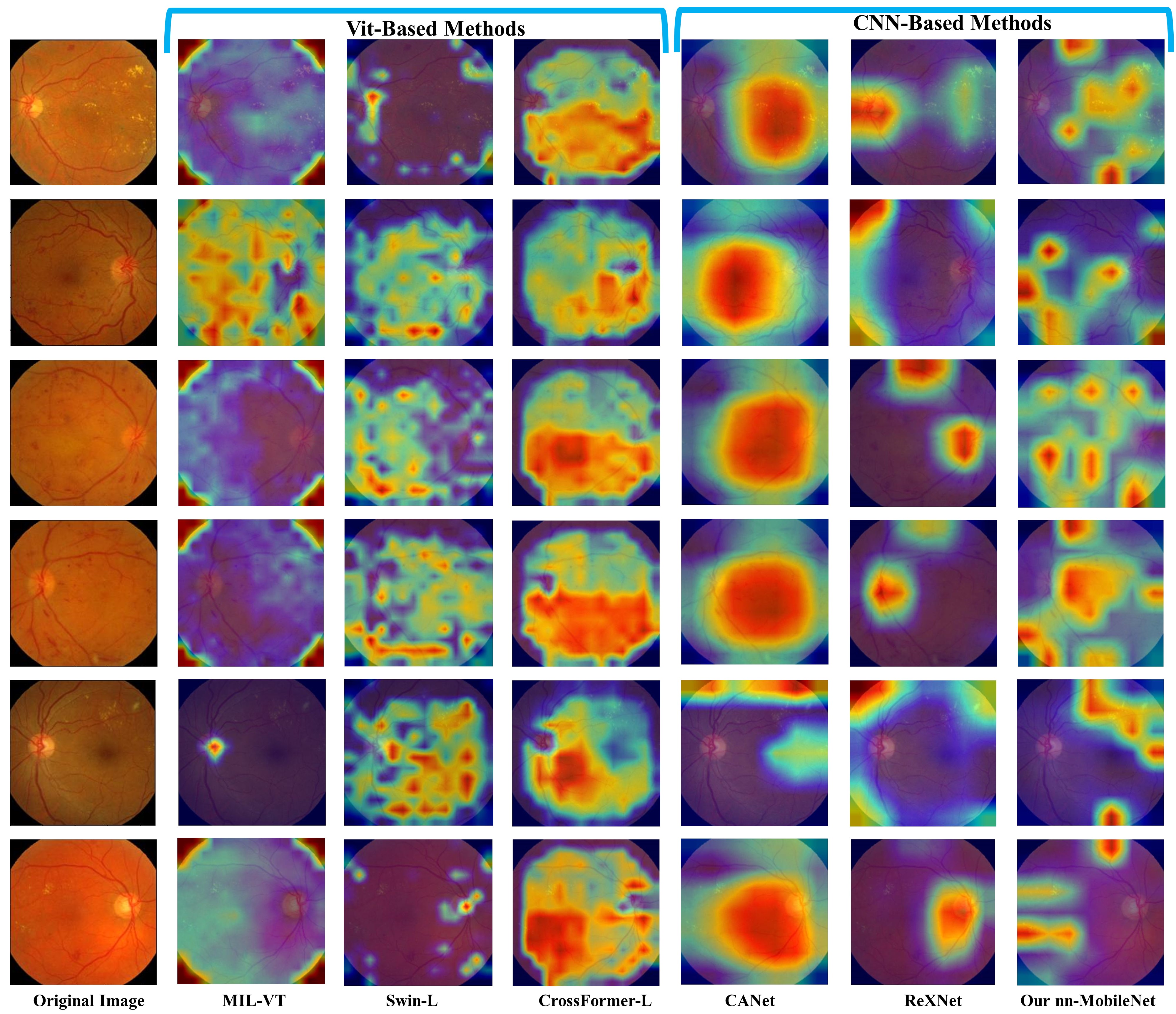}
    \caption{
The comparative visualization on the Messidor-1 dataset was performed utilizing CAM~\cite{selvaraju2017grad}.
We chose representative CNN/ViT-based methods with publicly available code, including MIL-VT~\cite{MIL-VT}, Swin-L~\cite{swin-L}, CrossFormer-L~\cite{crossformer}, CANet~\cite{CANet} and ReXNet~\cite{ReXNet}.
}
    \label{fig:cam}
\end{figure*}

\noindent \textbf{DR task performance.}
We compared the proposed method to a variety of existing state-of-the-art (SOTA) methods on three DR datasets (i.e., Messidor1, APOTS, and IDRID). The proposed method achieved the best performance on the IDRID dataset (AUC=91.6, ACC=73.1). Remarkably, the proposed method outperformed the best model (DETACH-DAW~\cite{DETACH}) by $8\%$ and $23.5\%$ in AUC and ACC (Table~\ref{tb:idrid}), respectively, on the IDRID dataset. We also found that the proposed method had the highest ACC and Kappa on the APOTS dataset with a similar AUC to the best-performed model (Table~\ref{tb:APOTS2019}). The proposed method achieved an equal performance to the best model (LAT~\cite{LAT}) on the referral DR classification task and the best performance on the normal DR classification task (Table~\ref{tb:messidor}). It is worth noting that most works (i.e., MIL-VT~\cite{MIL-VT}, LAT~\cite{LAT}, Zoom-in-Net~\cite{zoom-in-net}, Semi + Adv~\cite{sem+adv}, and CKML~\cite{vbxk-ckml}) were pre-trained on large-scale external datasets. Whereas, the proposed method was trained from scratch using the same benchmark datasets. 


\noindent \textbf{Multi-disease  abnormal detection performance.}
We also conducted experiments and comparisons to current SOTA methods on the multi-disease detection task. The proposed method achieved the best performance in terms of ACC and AUC, while the SatFormer-B~\cite{SatFormer-B} achieved the best performance in F1 (Table~\ref{tb:RFMiD}). However, our model (Param=34M) has fewer than half number of parameters of the SatFormer-B~\cite{SatFormer-B} (Param=78M). Even though the proposed model had a similar stem architecture to the RexNet, our heavy data augmentations and spatial dropout improved the ACC by $3.4\%$ and AUC by $4.4\%$.

\noindent \textbf{DME classification performance.}
The DME classification task was evaluated on the IDRID dataset following the protocol in~\cite{DETACH}. Table~\ref{tb:idrid} demonstrated that the proposed method surpassed the model with the best performance (DETACH+DAW~\cite{DETACH}) by $17.3\%$ on F1, $6.5\%$ on AUC, and $4.8\%$ on ACC. Compared to other SOTA methods (i.e., CANet~\cite{CANet}, Multi-task net~\cite{Multitask-Net}, MTMR-net~\cite{MTMR-NET}, and DETACH-DAW~\cite{DETACH}) that were jointly trained on multiple tasks, our proposed model was trained from scratch on the DME task only. 

\noindent \textbf{MICCAI MMAC 2023 Challenge.} 
Our well-calibrated nnMobileNet secured the third position in MICCAI MMAC 2023 Challenge~\cite{challenge} and was remarkably close to the top-ranking models (Table~\ref{tb:finalrank}). Whereas models that won the first and second places were ViT-based models pre-trined on large-scale external datasets using self-supervised learning. Consequently, their models were at least three times slower than ours' regarding the inference time on CPU (see Table~\ref{tb:finalrank}).

\section{Visual Interpretability}
We visualize the most discriminative regions of several representative methods using the gradient-weighted class activation map (Grad-CAM)~\cite{selvaraju2017grad} in the Messidor-1 dataset for the DR task. As shown in Fig.~\ref{fig:cam}, the proposed method showed the most accurate localization of diabetic lesions compared to the other baseline methods, e.g.. hard exudates, and hemorrhages. This observation aligns with our initial hypothesis that ViTs are typically employed to model the similarities between different patches. When dealing with small lesion blocks, localized lesions within many patches tend to be averaged out and overlooked, with ViTs favoring semantic comparisons between patches. Consequently, this leads to methods like Swin-L and CrossFormer producing CAM regions that are overly broad, hindering the precise localization of smaller lesions. It is noteworthy that MIL-VT compels each patch token to pass through a MIL (Multiple Instance Learning) head, essentially engaging in a pseudo-label learning process. We observed that this MIL attention mechanism tends to assign a uniform level of importance to all patch tokens, which disrupts the ability of ViT to learn the relationships between different patches. 
Compared with CNN methods, the multi-task network of CANet presents fitting challenges, indicating that despite the relatedness of the tasks, DME does not significantly enhance lesion localization in DR, possibly due to divergent interest patterns between the two tasks. Interestingly, The nn-mobilenet and ReXNet share the same model configuration, but the latter still struggles to accurately learn lesion representation. This situation underscores the importance of fine-tuning CNNs for improved performance. Finally, We observed that the ViT-based methods show inferior localization performance compared to CNN-based methods. However, other CNN-based baseline methods (i.e., ReXNet and CANet) only demonstrate coarse localization of the lesions. Whereas, the proposed method can accurately localize diabetic lesions. These findings suggest the importance of CNN in capturing small localized features for retinal disease diagnosis.

\section{Discussion and Conclusion}

In this article, we center our investigation on the question - \emph{Could CNN inherently be more suited to retinal disease (RD) tasks than ViTs ?} To address this, we embarked on a series of empirical studies, starting with fine-tuning a lightweight MobileNetV2. Through this process, we proposed a series of modifications to MobileNetV2, culminating in developing a tailored and lightweight model we denote as nnMobileNet. The proposed method surpasses ViT-based and multitask-driven models across various RD benchmarks. Remarkably, nnMobileNet achieves this superior performance without applying self-supervised pretraining on external datasets, highlighting the potential of CNNs in the domain of RD tasks.

In revisiting CNNs for RD tasks, we do not entirely negate the value of ViTs. It's evident from our findings that ViTs excel at capturing long-range dependencies better than CNNs. However, ViTs relying on extensive data for pre-training poses significant challenges for medical datasets subject to privacy concerns. Meanwhile, patterns of interest in natural images typically occupy a large portion of the image, and lesions in medical images often constitute a small fraction, making patch-based ViT relational understanding insufficient. Therefore, we offer the following recommendations for future model development in RD tasks: (i) CNNs are preferable in scenarios with limited retinal image data. (ii) CNNs have superior capabilities in capturing fine-grained local features, particularly for RD tasks focused on small lesions. (iii) Integrating CNNs with ViTs could be a viable solution. (iv) Emphasize data characteristics and model fine-tuning. (v) Large-kernel convolutions could address limitations in capturing long-range dependencies.
In the end, we believe that results will challenge several widely held views and prompt people to rethink the importance of convolution in RD.

\noindent\textbf{ACKNOWLEDGMENTS} The work was partially supported by NIH (R01EY032125, R01DE030286) and the state of Arizona via the Arizona Alzheimer’s Consortium, United States.

{
    \small
    \bibliographystyle{ieeenat_fullname}
    \bibliography{main}
}


\end{document}